\documentclass[aps,pre,twocolumn,groupedaddress,showpacs]{revtex4}

\usepackage{amsmath}
\usepackage{amssymb}
\usepackage{epsfig}
\newcommand {\dr}{{\mathrm d}\mathbf{r}}

\newcommand {\dk}{{\mathrm d}\mathbf{k}}
\newcommand {\dd}{{\mathrm d}}
\newcommand {\rr}{\mathbf{r}}
\newcommand {\RR}{\mathbf{R}}
\newcommand {\kk}{\mathbf{k}}
\newcommand {\etal}{\begin{itshape}et al\end{itshape}.}

\newcommand {\erfc}{{\mathrm e}{\mathrm r}{\mathrm f}{\mathrm c}}
\begin{document}


\title{Density functional theory for the freezing of soft--core fluids}


\author{A.J. Archer}
\email{Andrew.Archer@bristol.ac.uk}
\affiliation{H.H. Wills Physics Laboratory,
University of Bristol, Bristol BS8 1TL, United Kingdom}


\date{\today}

\begin{abstract}
We present a simple density functional theory for the solid phases of systems of
particles interacting via soft--core potentials. In particular, we apply the
theory to particles interacting via repulsive point Yukawa and Gaussian pair
potentials. We find qualitative agreement with the established phase
diagrams for these systems. The theory is able to account for the bcc--fcc solid
transitions of both systems and the re--entrant melting that the Gaussian system
exhibits.
\end{abstract}


\maketitle

\section{Introduction}

When an ensemble of particles is at a sufficiently high density, the fluid
will often freeze to form a crystalline solid. This freezing is driven by the
repulsion that invariably exists between particles when at close separations.
The question that naturally arises is: how much repulsion is needed for
freezing and what crystal structures are formed? The answer is that even
particles interacting via repulsive
pair potentials $v(r)$ that are finite for all separations
$r$ can freeze \cite{Likos,LikosetalPRE1998,paper4b}. The particular crystal
structure that is
formed depends upon the form of the particle interactions and especially on the
`softness' of the decay of $v(r)$ as $r \rightarrow \infty$
\cite{Agrawal:KofkePRL1995}. Furthermore, there can
be solid--solid transitions, with different crystal structures being stable
in different portions of the phase diagram. In this paper we present
a simple
density functional theory (DFT) \cite{Bob} for determining the location of the
melting and solid--solid phase boundaries for soft--core particles.

For the purposes of this paper, we define `soft--core' particles as those with
purely repulsive pair potentials $v(r)$, for which the integral over all space
of $v(r)$ is finite, i.e.\ $\int \dr v(r) < \infty$. Alternatively, one can
define soft--core potentials as those for which the Fourier transform
$\hat{v}(k)$ of the pair potential exists. Commonly encountered
(model) potentials in the theory of liquids such as the Coulomb potential,
the hard--sphere potential or
the Lennard--Jones potential \cite{HM} do not fall into this category.
However, a wide class of fluids can be modelled by particles interacting via
potentials that do fall into the soft--core category. One common example is the
Yukawa core model (YCM):
\begin{equation}
v(r) \, =\, \frac{\epsilon \exp(-\lambda r)}{\lambda r},
\label{eq:pairpot_Yuk}
\end{equation}
where $\lambda>0$ and $\epsilon>0$. Such a potential is used to model the
effective interaction between charged point particles,
where the Coulomb interaction between the particles is screened by a background
medium. The effects of the screening are incorporated in the parameter
$\lambda$ \cite{levin2002,HansenLowen99}. Examples of systems where the particle
interactions can be modelled by a Yukawa pair potential range from charged
colloidal solutions \cite{HansenLowen99,hynninen:dijkstra2003,kleinetalJPCM2002}
to dusty plasmas
\cite{piel:melzer2002,Valuina,rosenfeldPRE1994,farouki:hamaguchi1994}.

Another soft--core potential is the Gaussian core model (GCM):
\begin{equation}
v(r) \, =\, \epsilon \exp (-\lambda^2 r^2).
\label{eq:pairpot_GCM}
\end{equation}
The freezing behaviour of a fluid composed of particles interacting via such a
potential aroused much interest due to the novel re--entrant melting behaviour:
for certain temperatures, on increasing the fluid density, the fluid freezes.
However, on further increasing the density, the crystal
re--melts \cite{Likos}. The high density phase of the GCM
is the fluid state \cite{Likos, Stillinger4, paper3,paper5,paper4a,
PrestipinoetalPRE2005}. A Gaussian potential is used to model
the effective interaction between the centres of mass of
polymers, star--polymers and dendrimers in solution \cite{Likos,IngoJCP}. In
this case $\lambda^{-1} \simeq R$, the radius of gyration of the polymers.

The DFT theory for freezing presented here is a simple qualitatively accurate
theory which, for the cases we have tested, gives the correct topology of the
phase diagram, including the existence of solid--solid phase transitions --
i.e. the theory
incorporates in a simple way much of the physics of the solid phases of soft
core particles. Furthermore, we believe the present theory is of general
interest to
the classical DFT community, since many DFT theories are unable to describe
solid--solid coexistence. For example, in Refs.\
\cite{LairdKrollPRA1990,McConnellGastPRE1996} the authors applied DFTs that
are very successful for hard--spheres to fluids composed of particles
interacting via Yukawa and inverse
power pair potentials and found them not to predict the bcc phase. The
present work may give some insight into what is required in a DFT in order to
describe a solid--solid transition -- see also Ref.\ \cite{GrohSchmidtJCP2001}.

This paper is laid out as follows: In Sec.\ \ref{sec:2} we describe the DFT
theory. In Sec.\ \ref{sec:3} we apply the theory to the GCM and then in Sec.\
\ref{sec:4} to the YCM.
Finally, in Sec.\ \ref{sec:5} we discuss our results and draw some conclusions.

\section{The DFT theory}
\label{sec:2}

Given an expression for the Helmholtz free energy of a system, one can obtain
all other thermodynamic quantities. It can be shown that the Helmholtz
free energy ${\cal F}$ is a unique functional of the one body density profile of
the system, $\rho(\rr)$ \cite{Bob}. We can divide the Helmholtz
free energy into two parts:
\begin{equation}
{\cal F}[\rho] = {\cal F}_{id}[\rho] + {\cal F}_{ex}[\rho].
\label{eq:F}
\end{equation}
The first term is the ideal gas contribution \cite{Bob}:
\begin{equation}
{\cal F}_{id}[\rho] = k_BT \int\dr\, \rho(\rr)[\ln (\rho(\rr)\Lambda^3)-1],
\label{eq:F_id}
\end{equation}
where $T$ is the temperature, $k_B$ is Boltzmann's constant and $\Lambda$ is the
thermal de--Broglie wavelength. The excess part is given formally by \cite{Bob}:
\begin{equation}
{\cal F}_{ex}[\rho] = \frac{1}{2} \int\dr\int\dr'\, v(\rr,\rr')
\rho(\rr)\rho(\rr') \int_0^1 \dd a g(\rr,\rr';a)
\label{eq:F_ex}
\end{equation}
where $g(\rr,\rr';a)$ is the inhomogeneous radial distribution function
corresponding to a system of particles interacting via the pair potential
$a v(\rr,\rr')$, i.e.\ Eq.\ (\ref{eq:F_ex}) is formally derived by `turning--on'
the interactions between the particles (via the parameter $0\leq a \leq 1$) and
integrating $g(\rr,\rr';a)$ as $a$ is increased from 0 to 1 keeping $\rho(\rr)$
fixed \cite{Bob}. The main
approximation in our theory involves replacing the function
$\int_0^1 \dd a g(\rr,\rr';a)$ by
a simple ansatz. This is given below; first we make a few remarks about
$\rho(\rr)$.

When in the uniform fluid state, the one-body density is a constant,
$\rho(\rr)=\rho \equiv N/V$, the average number density, where $N$ is the number
of particles and $V$ is the volume of the system. However, when the system
freezes into a solid the density becomes periodic, i.e.\ the symmetry breaks
and $\rho(\rr)=\rho(\rr-\RR_i)$, where $\RR_i$ is a lattice vector for the solid
phase. An approximation that is often made in DFT studies of freezing
\cite{Likos,BausJCP1990}, is to assume that the density profile of the
solid is made up of Gaussian peaks centred on each lattice site:
\begin{equation}
\rho(\rr) = \sum_{i=1}^N G(\rr-\RR_i)
\label{eq:rho_solid}
\end{equation}
with
\begin{equation}
G(\rr-\RR_i)= \left(\frac{\alpha}{\pi} \right)^{3/2}
\exp[-\alpha (\rr-\RR_i)^2],
\label{eq:rho_solid_G}
\end{equation}
where $\alpha$ is a parameter which describes how localised the particles are
around each lattice site. This density profile assumes the normalisation
condition that there is one particle per lattice site. Of course, this
assumption need not necessarily be
true. We can expect to find vacancies in the crystal, and indeed for the present
soft--core systems we may find another type of defect: double occupancy,
since the
particle cores can overlap. However, we expect the proportion of defects to
be small, and assume a perfect crystal with sites singly occupied.

Together with the assumption that $\rho(\rr)$ takes the form in Eq.\
(\ref{eq:rho_solid}) we make the further assumption that
${\cal F}_{ex}$ takes the form:
\begin{equation}
{\cal F}_{ex} = \frac{1}{2} \sum_{i \neq j} \int\dr\int\dr'\, v(\rr-\rr')
G(\rr-\RR_i)G(\rr'-\RR_j).
\label{eq:F_ex_2}
\end{equation}
This constitutes an RPA like approximation for
$\int_0^1 \dd a g(\rr,\rr';a)$ in Eq.\ (\ref{eq:F_ex}), with the
`self--energy' term in the summation over lattice vectors, $\RR_i=\RR_j$,
being subtracted. When $\alpha$ is sufficiently large so that there is
negligible overlap between the density peaks on neighbouring lattice sites, our
assumption is equivalent to:
\begin{equation}
\int_0^1 \dd a g(\rr,\rr';a)=
\begin{cases}
0 \hspace{5mm} |\rr-\rr'| < l \\
1 \hspace{5mm} |\rr-\rr'| > l,
\end{cases}
\label{eq:g_approx}
\end{equation}
where the length $l \sim b_1/2$ and $b_1$ is the distance between nearest
neighbour lattice sites. This
approximation therefore constitutes quite a drastic simplification of the
function $\int_0^1 \dd a g(\rr,\rr';a)$, which neglects much of the
information about correlations in the system that this function contains.

Eq.\ (\ref{eq:F_ex_2}) is very appealing because it takes the
form of a double convolution and can therefore be written in the form:
\begin{equation}
{\cal F}_{ex}(\rho,\alpha) = \frac{1}{2} \sum_{i \neq j} \frac{1}{(2 \pi)^3}
\int\dk \exp(i \kk \cdot \RR_{ij}) \hat{v}(k) \hat{G}(k) \hat{G}(k)
\label{eq:F_ex_3}
\end{equation}
where $\RR_{ij}=\RR_j-\RR_i$ and $\hat{G}(k)$ is the Fourier transform of
$G(r)$. Since $G(r)$ is a Gaussian function, then so is $\hat{G}(k)$.

The ideal gas contribution to the Helmholtz free energy (\ref{eq:F_id}) also
takes a simple form if we assume $\rho(\rr)$ is given by Eq.\
(\ref{eq:rho_solid}) and we further assume that $\alpha$
takes values sufficiently large that the
overlap between the Gaussian density peaks on neighbouring lattice sites is
negligible. Then the ideal gas part of the Helmholtz free energy,
Eq.\ (\ref{eq:F_id}), is simply (see e.g.~\cite{LairdKrollPRA1990})
\begin{equation}
{\cal F}_{id}(T,N,\alpha) = N k_BT
\left[ \frac{3}{2}\ln \left( \frac{\Lambda^2 \alpha}{\pi} \right)
\,-\, \frac{5}{2} \right].
\label{eq:F_ideal_crystal}
\end{equation}

Given a pair potential $v(r)$, for which the Fourier transform $\hat{v}(k)$
exists, Eqs.\ (\ref{eq:F_ex_3}) and (\ref{eq:F_ideal_crystal}) together provide
an expression for the
free energy ${\cal F}$ which is a function of temperature, the average
density $\rho$, the parameter $\alpha$ and the set of lattice vectors
$\{\RR_i\}$. For a given state point $(T,\rho)$ and lattice structure,
we assume that the parameter $\alpha$ is determined by the minimisation
condition $(\partial {\cal F}/\partial \alpha)_{\alpha=\alpha_{min}}=0$
\cite{footnote} and we assume that the Helmholtz free energy $F={\cal
F}(\alpha_{min})$. We can therefore calculate the Helmholtz free energy for a
number of candidate crystal structures and then the equilibrium crystal
structure is that with the lowest free energy. In order
to determine the melting phase boundary we could compare this minimal value of
$f=F/V$ with that calculated from the theory applied
to the liquid state and then perform the common tangent construction
between these free energies (which is equivalent to equating chemical potentials
and pressures in the coexisting phases \cite{Likos}). In the liquid state, where
$\rho(\rr)=\rho$, the ideal gas contribution to the Helmholtz
free energy (\ref{eq:F_id}) becomes
\begin{equation}
{\cal F}_{id}(\rho) = Nk_BT[\ln (\rho \Lambda^3)-1],
\label{eq:F_id_liq}
\end{equation}
and the excess contribution, obtained from Eqs.\ (\ref{eq:F_ex}) and
(\ref{eq:g_approx}), is:
\begin{equation}
{\cal F}_{ex}^{liq}(\rho) = 2 \pi N \rho \int_l^{\infty} \dd r r^2 v(r).
\label{eq:F_ex_liq}
\end{equation}
The latter constitutes a very crude approximation for the liquid state free
energy. Given that our aim here is to construct a theory which is above all
simple, but which is
still able to provide a qualitative description of the solid phases of
soft--core particles, we choose to employ a Lindemann
criterion \cite{Lindemann} to calculate the solid
melting curves, rather than compare our solid free energy with that of the
liquid. The Lindemann criterion simply states that when the
root--mean--square displacement, $\sigma \equiv (\left< r^2 \right> - \left< r
\right>^2)^{1/2}$, of a particle about its equilibrium position is roughly 10\%
of the nearest neighbour distance $b_1$, the crystal will melt.
For the Gaussian density profile, Eq.\ (\ref{eq:rho_solid}),
$\sigma = \sqrt{3/2 \alpha}$ and we determine approximate melting
boundaries from the locus defined by $\sigma/b_1=0.1$.

One can improve upon Eq.\ (\ref{eq:rho_solid}) as an approximation for the
density profile in the crystal: in order to incorporate the effects of
anisotropy in the density peaks around each lattice site one can assume the
density profile is of the following form:
\begin{eqnarray}
\rho(\rr)=\sum_{i=1}^N \left(\frac{\alpha}{\pi} \right)^{3/2}
{\mathrm e}^{-\alpha (\rr-\RR_i)^2}[1+\tau \alpha^2 f(\rr-\RR_i)],
\label{eq:rho_solid_2}
\end{eqnarray}
where the function $f(\rr) =x^4+y^4+z^4-3r^4/5$ is the leading term for the unit
cell anisotropy in cubic lattices
\cite{VondeLageandBethePR1947,TarazonaPRL2000} and $\tau$ is an
anisotropy parameter. One then minimises the free energy with respect to
$\tau$, as well as $\alpha$. We attempted such an approach for the GCM, but we
found it made no significant change to the phase diagram that we obtained from
the simple choice (\ref{eq:rho_solid}).

\section{Application to the GCM}
\label{sec:3}

\begin{figure}
\includegraphics[width=8cm]{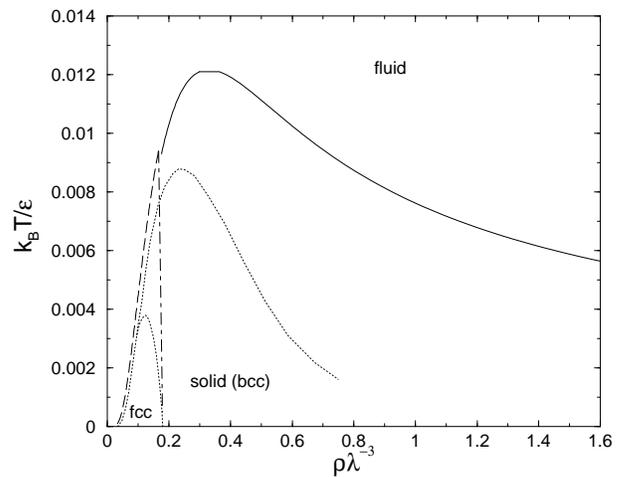}
\caption{Phase diagram of the GCM. The solid line is the bcc melting boundary
and the dashed line is the fcc melting boundary, both determined
using the Lindemann criterion -- see the text. The dot--dashed line is
the locus of points where the Helmholtz free energies of the bcc and fcc phases
are equal. The dotted lines are the phase boundaries obtained from the computer
simulations of Prestipino {\it et al.}~\cite{PrestipinoetalPRE2005}.}
\label{fig:1}
\end{figure}

In the GCM, the interparticle pair potentials are given by Eq.\
(\ref{eq:pairpot_GCM}). The Fourier transform of this Gaussian potential is also
a Gaussian and so the excess Helmholtz free energy given by Eq.\
(\ref{eq:F_ex_3}) takes the particularly simple form:
\begin{equation}
{\cal F}_{ex}^{GCM} = \sum_{i \neq j} \frac{\epsilon \gamma^{3/2}}{2 \lambda^3}
\exp(-\gamma
R_{ij}^2),
\label{eq:F_GCM}
\end{equation}
where $\gamma=(1/\lambda^2+2/\alpha)^{-1}$ and $R_{ij}=|\RR_{ij}|$.
Using Eqs.\ (\ref{eq:F_ideal_crystal}) and (\ref{eq:F_GCM}) as our approximation
for the Helmholtz free energy of the crystal, we calculate the phase
diagram for the GCM. The results are displayed in Fig.\ \ref{fig:1}.
Recently, Prestipino \etal\ \cite{PrestipinoetalPRE2005}
made an accurate determination of the GCM phase
diagram using Monte Carlo simulations, so we are able to compare with these
essentially exact results (see also Ref.\ \cite{paper4a}).
We find, as did Prestipino \etal\ \cite{PrestipinoetalPRE2005}, that at low
temperatures, on increasing the density, the fluid first freezes to form a
face--centred--cubic (fcc) crystal, and then on further increasing the density
there is a transition from the fcc to a body--centred--cubic (bcc) crystal
\cite{footnote1} -- see Fig.\ \ref{fig:1}.
Performing the common tangent construction between the bcc and fcc free
energies, we find that the two-phase region between the two crystal phases is
very narrow, the difference between the coexisting densities $\Delta \rho
\lambda^{-3} \sim 10^{-4}$ -- see also the inset to Fig.\ 9 of Ref.\
\cite{paper4a}. Since we are mostly interested in providing a simple theory
which accounts for the topology of the phase diagram, we
determined the density at which the Helmholtz free energy of the bcc equals that
of the fcc structure for a given temperature \cite{footnote2}.
The resulting line is plotted in
Fig.\ \ref{fig:1}. The present theory also accounts for the most striking
feature of the GCM phase diagram: the re--entrant melting of the bcc phase --
i.e.\ for a given (low) temperature, on increasing the fluid density it freezes,
but on further increasing the density the crystal remelts. The high density
phase of the GCM is a fluid. This means there is also a maximum temperature for
which there is crystal. The present theory predicts this maximum
to be at a temperature $k_BT/\epsilon \simeq 0.012$, whereas it is actually at
 $k_BT/\epsilon \simeq 0.009$ \cite{PrestipinoetalPRE2005}. In general, the
 present theory over estimates the region of stability for the solid phases.

\begin{figure}
\includegraphics[width=8cm]{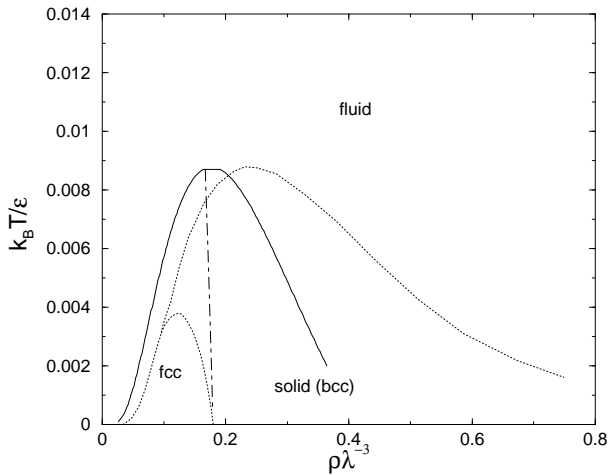}
\caption{Same as Fig.~\ref{fig:1}, except now the solid line is the
locus of points where the Helmholtz free energies of the bcc and liquid phases
are equal. The liquid state free energy is calculated using the rather crude
approximation in Eq.\ (\ref{eq:F_ex_liq}).}
\label{fig:1_a}
\end{figure}

We also determine an approximate melting boundary by calculating the
locus in the phase diagram where the bcc free energy equals the
liquid state free energy \cite{footnote2},
where the liquid Helmholtz free energy is
calculated using the crude approximation, Eq.\ (\ref{eq:F_ex_liq}), which for
the GCM becomes:
\begin{equation}
{\cal F}_{ex}^{GCM,liq} = \frac{1}{2} N \rho \pi^{3/2} \epsilon \lambda^{-3}
\left[ \frac{2 \lambda l}{\sqrt{\pi}} \exp(-\lambda^2l^2)+\erfc(\lambda l)
\right],
\label{eq:F_GCM_liq}
\end{equation}
where $\erfc(x)=1-2 \pi^{-1/2}\int_0^x {\mathrm d} t \exp(-t^2)$ is the
complimentary error function. Whilst we know $l \simeq b_1/2$, there is no
constraint on a particular value. We choose the value
$l \simeq 0.58(2/\rho)^{1/3}$ (recall that for the bcc crystal
$b_1=(\sqrt{3}/2)(2/\rho)^{1/3}$). This value of $l$ is chosen so
that the maximum temperature that freezing occurs is predicted to be roughly at
the same temperature as that from the results of Prestipino \etal\
\cite{PrestipinoetalPRE2005}. Note, on the scale of Fig.\ \ref{fig:1_a},
using this approach there is no difference between the predicted locations of
the bcc--liquid melting boundary and the fcc--liquid melting boundary.
Given the crude nature of the theory for the
liquid, the results are surprisingly good -- see Fig.\ \ref{fig:1_a}.

The main feature of the GCM phase diagram that is not accounted for by the
present theory, and which can be seen in the results of Prestipino \etal\
\cite{PrestipinoetalPRE2005} (see also Fig.\ \ref{fig:1}),
is that the transition line between the fcc and
bcc phases is not the (almost) straight line predicted by the present theory.
Instead, on following the fcc--bcc boundary as the temperature is increased, the
boundary obtained in simulation
curves over to lower densities and in fact reaches a maximum and then
bends down to lower temperatures and densities, such that there is a small
window in the phase diagram where for
temperatures around $k_BT/\epsilon \simeq 0.0035$, on increasing the density,
the fluid first freezes to form the bcc, then the fcc and then the bcc phase
again, before finally re--melting \cite{PrestipinoetalPRE2005}.
This bending over of the fcc--bcc boundary was ascribed to anharmonic effects in
the crystal \cite{PrestipinoetalPRE2005}. In an attempt to incorporate this
effect we assumed the
density took the form given in Eq.\ (\ref{eq:rho_solid_2}). Within the
present theory anharmonic effects do indeed become more
significant in the high temperature part of the fcc portion of the phase
diagram. However, they do not become significant
enough to result in the fcc--bcc boundary curving over, as in the results of
Prestipino \etal\ \cite{PrestipinoetalPRE2005}. Assuming
(\ref{eq:rho_solid_2}) leaves the phase diagram unchanged when plotted on the
scale given in Fig.\ \ref{fig:1}.

For the results presented in Fig.\ \ref{fig:1} we summed over 40 shells of
lattice vectors (including any more does not change the value calculated for the
free energy). However, it is interesting to note that if one includes only the
first two shells -- i.e.\ nearest neighbour and next nearest neighbour
contributions only, then the results are qualitatively unchanged
\cite{myThesis}. This is
not too surprising, given the short ranged nature of the GCM pair potential.

\section{Application to the YCM}
\label{sec:4}

\begin{figure}
\includegraphics[width=8cm]{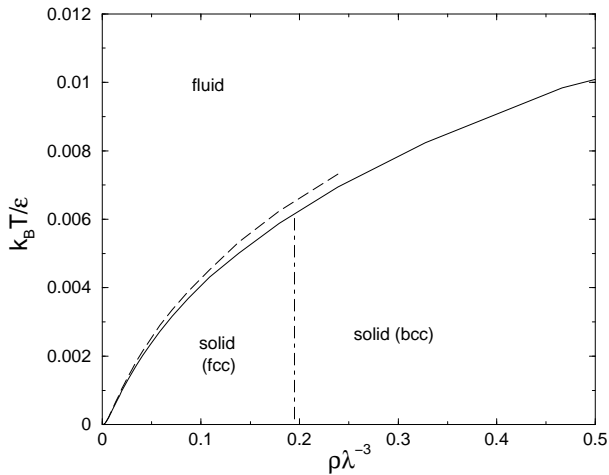}
\caption{Phase diagram of the YCM. The solid line is the bcc melting boundary
and the dashed line is the fcc melting boundary, both determined
using the Lindemann criterion -- see the text. The dot--dashed line is
the locus of points where the Helmholtz free energies of the bcc and fcc phases
are equal.}
\label{fig:2}
\end{figure}

The Fourier transform of the YCM pair potential, Eq.\ (\ref{eq:pairpot_Yuk}), is
$\hat{v}(k)=4 \pi \epsilon \lambda^{-1}(\lambda^2+k^2)^{-1}$. Using this,
together with Eq.\ (\ref{eq:F_ex_3}), we obtain the following expression for
the YCM excess Helmholtz free energy:
\begin{eqnarray} \notag
{\cal F}_{ex}^{YCM} = \sum_{i \neq j}
\frac{\epsilon {\mathrm e}^{\lambda^2/2 \alpha}}{4 \lambda R_{ij}} \Bigg[
{\mathrm e}^{-\lambda R_{ij}}
\erfc \left( \frac{\lambda}{\sqrt{2 \alpha}}-R_{ij}
\sqrt{\frac{\alpha}{2}} \right) \\
-{\mathrm e}^{\lambda R_{ij}}
\erfc \left( \frac{\lambda}{\sqrt{2 \alpha}}+R_{ij}
\sqrt{\frac{\alpha}{2}} \right) \Bigg].
\label{eq:F_YCM}
\end{eqnarray}
This approximation, together with Eq.\ (\ref{eq:F_ideal_crystal}), is our
expression for the Helmholtz free energy of the solid phases of the YCM.
For a given state point $(\rho,T)$ and set of lattice vectors $\{\RR_i \}$,
we minimise the free energy with respect to the parameter
$\alpha$. We estimate the melting boundaries using the Lindemann criterion --
i.e.\ the locus defined by $\sigma/b_1=0.1$. In Fig.\ \ref{fig:2} we display
the resulting phase diagram. For the YCM, the crude estimate for the liquid
state excess Helmholtz free energy, Eq.\ (\ref{eq:F_ex_liq}), does not give
physically acceptable results. For some choices of $l$ in (\ref{eq:F_ex_liq}) it
(incorrectly) predicts that the YCM exhibits re--entrant melting. We therefore
employ only the Lindemann criterion for determining melting boundaries in this
section.

\begin{figure}
\includegraphics[width=8cm]{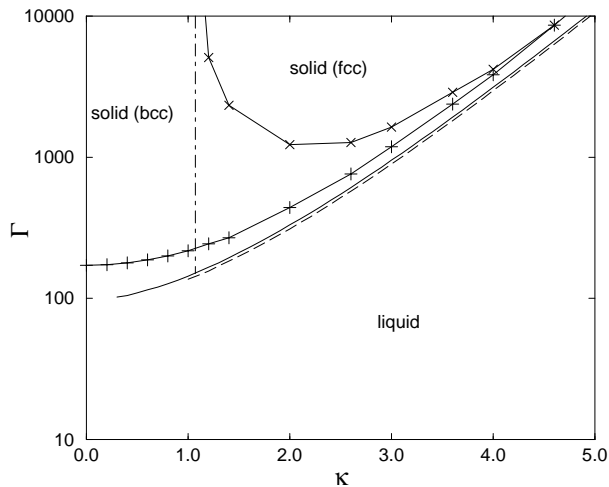}
\caption{Same as Fig.\ \ref{fig:2}, except here the results are plotted in terms
of the variables $\kappa$ and $\Gamma$. The symbols joined by solid lines are
the simulation results of Hamaguchi {\it et al.}~\cite{HamaguchietalPRE1997}.}
\label{fig:3}
\end{figure}

The low density portion of the YCM phase diagram is qualitatively similar to
that of the GCM -- i.e.\ for sufficiently low temperatures, on increasing the
density the fluid first freezes to
form an fcc crystal, then, at higher densities, there is a transition to the
bcc. The biggest difference between the YCM and the GCM phase diagram is that
there is no re-entrant melting in the YCM. This is because the divergence of
the YCM pair potential as $r \rightarrow 0$ means that the particles behave more
and more like hard spheres as the density is increased, where the effective hard
sphere diameter depends upon the state point $(T,\rho)$, so that there is
freezing of the YCM at all temperatures. For the GCM this is not the case. Note
that a divergence in the pair potential at $r \rightarrow 0$ does not
automatically mean that there is no re-entrant melting
\cite{WatzlaweketalPRL1999}.

Just as for the GCM, the present theory for the YCM
also fails to account for the curving over to lower densities of the fcc-bcc
boundary as the temperature is increased in the YCM. This can be seen in Fig.\
\ref{fig:3}, where we plot the YCM phase diagram that we obtain together with
that of Hamaguchi \etal\ \cite{HamaguchietalPRE1997} (see also
Refs.\ \cite{kremeretalPRL1986, robbinsetalJCP1988, rosenberg:thirumalaiPRA1987,
meijer:frenkelJCP1991, HamaguchietalJCP1996, Hoy:RobbinsPRE2004, Hopkins1}),
plotted in terms of the variables $\Gamma=\beta \epsilon/\lambda a$ and
$\kappa=\lambda a$, where $a=(3/4 \pi \rho)^{1/3}$. These are commonly
considered variables when using the Yukawa potential (\ref{eq:pairpot_Yuk}) to
model the interactions in plasma systems. We see that the present simple
theory is able to account qualitatively for the YCM phase diagram.

\section{Discussion and conclusions}
\label{sec:5}

We have constructed a simple DFT for the solid phases of soft--core
particles. The theory is able to account for the transition from a fcc to
a bcc solid and also able to determine whether the system exhibits re-entrant
melting, as is the case for the GCM, or not, as in the case for the YCM (when a
Lindemann criterion is used to determine the melting boundaries).
This makes the DFT useful, since many DFTs are not able to
describe solid--solid coexistence in soft core fluids
\cite{LairdKrollPRA1990,McConnellGastPRE1996}. There
are some DFT theories able to describe solid--solid transitions in other
(hard--core) model fluids -- see for example
Refs.\ \cite{Likos, LikosetalJPCM1994, RasconetalJCP1997}. In fact, the present
theory bears some similarities in its structure to that of Likos \etal\
\cite{Likos,LikosetalJPCM1994}. This can be traced to the use of the
Gibbs--Bogoliubov inequality to construct their theory. When one applies this
inequality, one obtains the following equation (Eq.\ (4.18) in Ref.\
\cite{Likos}):
\begin{equation}
{\cal F}[\rho] \leq {\cal F}_0[\rho]+\frac{1}{2} \int \dr \int \dr'
g_0(\rr,\rr') \rho(\rr) \rho(\rr') v(\rr-\rr'),
\label{eq:G-B}
\end{equation}
where $g_0$ and ${\cal F}_0$ are the pair distribution function and the
Helmholtz free energy respectively of the reference system; Likos \etal\
\cite{Likos,LikosetalJPCM1994} used a hard--sphere fluid as the reference
system. If we compare Eq.\ (\ref{eq:G-B})
with Eq.\ (\ref{eq:F_ex}) we can see that depending on our approximations for
$g_0(\rr,\rr')$ in (\ref{eq:G-B}) and $\int_0^1 \dd a g(\rr,\rr';a)$ in
(\ref{eq:F_ex}), one can end up with theories that have a similar structure.

One also sees similar features when we compare our expression for the GCM
Helmholtz free
energy, Eqs.\ (\ref{eq:F_ideal_crystal}) and (\ref{eq:F_GCM}), with that
obtained by Lang \etal\ \cite{paper4a,Likos} using the Gibbs-Bogoliubov
inequality together with the Einstein model as their reference system. Their
resulting Helmholtz
free energy is almost exactly the same as that in the present theory
(the free energies differ only by $k_BT$ per particle,
with that of Lang \etal\ being
lower). In the theory of Lang \etal\ \cite{paper4a,Likos} the Einstein model
spring constant is the variational parameter for minimising the free energy,
whereas in the present theory it is the parameter $\alpha$ in Eq.\
(\ref{eq:rho_solid}). However, formally these parameters play exactly the same
role. Use of the Gibbs-Bogoliubov inequality therefore seems to lead to
theories with a structure similar to the theory
presented here. When taking our present approach, i.e.~using the Lindemann
criterion to determine the melting boundaries, the difference of $Nk_BT$ between
the Helmholtz free energy of the present theory and
that of Lang \etal\ \cite{paper4a} makes
no difference since this term is independent of $\alpha$. However, it would
matter if we were to compare our result for the solid free energy with that
obtained for the liquid from some other theory, more accurate than that from
Eq.\ (\ref{eq:F_ex_liq}).

Some soft core fluids exhibit freezing to states with multiple occupancies of
each lattice site \cite{LikosetalPRE1998,Archer6}. In order to apply the present
theory to such systems, some modification of the theory is required. Firstly,
one must assume a generalisation of Eq.\ (\ref{eq:rho_solid}) for the
density profile of the crystal:
\begin{equation}
\rho(\rr) = \eta \sum_{i=1}^N \left(\frac{\alpha}{\pi} \right)^{3/2}
{\mathrm e}^{-\alpha (\rr-\RR_i)^2},
\label{eq:rho_solid_eta}
\end{equation}
where $\eta$ is the average lattice site occupancy.
$\eta$ should be treated as a
parameter to minimise the Helmholtz free energy, in the same way as with the
parameter $\alpha$. In this case, there would be two minimisation conditions to
be satisfied: $(\partial {\cal F}/\partial \alpha)_{\alpha=\alpha_{min}}=0$
{\em and} $(\partial {\cal F}/\partial \eta)_{\eta=\eta_{min}}=0$. One
would then assume that the Helmholtz free energy $F={\cal
F}(\alpha_{min},\eta_{min})$. This would also be the scheme to apply if one
intended to study the effect of lattice defects in the present YCM and GCM
systems. However, in these cases one would expect $\eta \simeq 1$. For systems
exhibiting multiple occupancies of each lattice site, we expect one
would also have to make a different approximation for the function
$\int_0^1 \dd a g(\rr,\rr';a)$ in Eq.\ (\ref{eq:F_ex}). We would propose the
following generalisation of Eq.\ (\ref{eq:F_ex_2}):
\begin{equation}
{\cal F}_{ex}[\rho] =\frac{\eta}{2}\sum_{j=1}^N \sum_{i=1}^N (\eta-\delta_{ij})
\int\dr\int\dr' v(\rr,\rr')
G(\rr-\RR_i)G(\rr'-\RR_j).
\label{eq:g_ansatz_2}
\end{equation}

From a more general point of view, the present theory seems to provide a good
qualitative description of the solid phases of soft--core systems. The Yukawa
potential, (\ref{eq:pairpot_Yuk}), is used to model the effective interaction
between charged colloidal particles \cite{levin2002, HansenLowen99,
hynninen:dijkstra2003, kleinetalJPCM2002}. For example, the phase diagram of
polystyrene particles suspended in a potassium chloride solution can be mapped
on to that of the YCM \cite{Pusey}. The present theory should also be relevant
to soft matter systems, for example polymeric micelles
\cite{McConnellGastPRE1996}, star polymers \cite{WatzlaweketalPRL1999} and
dendrimers \cite{dendrimers}. Given an effective pair potential $v(r)$ between
such objects \cite{Likos}, one could use the present theory to calculate an
approximate phase diagram.

\section*{Acknowledgements}
I thank Christos Likos and Bob Evans for useful discussions and for critically
reading this manuscript. I gratefully
acknowledge the support of EPSRC under grant number GR/S28631/01.


\begin{thebibliography}{99}

\bibitem{Likos}
C.N. Likos, Phys. Reports, {\bf 348}, 267 (2001).

\bibitem{LikosetalPRE1998}
C.N. Likos, M. Watzlawek, and H. L\"owen, Phys. Rev. E {\bf 58}, 3135 (1998).

\bibitem{paper4b}
C.N. Likos and A. Lang and M. Watzlawek and H. L{\"o}wen, Phys. Rev. E, {\bf
63}, 031206 (2001).

\bibitem{Agrawal:KofkePRL1995}
R. Agrawal and D.A. Kofke, Phys. Rev. Lett, {\bf 74}, 122 (1995).

\bibitem{Bob}
R. Evans, in {\it Fundamentals of Inhomogeneous Fluids},
ed. D. Henderson, (Dekker, New York), (1992), ch. 3.

\bibitem{HM}
J-P. Hansen and I.R. McDonald, {\it Theory of Simple Liquids}, Academic,
London, (1986), 2nd ed.

\bibitem{levin2002}
Y. Levin, Rep. Prog. Phys., {\bf 65}, 1577 (2002).

\bibitem{HansenLowen99}
J-P. Hansen and H. L\"owen, Annu. Rev. Phys. Chem., {\bf 51}, 209 (2000).

\bibitem{hynninen:dijkstra2003}
A.P. Hynninen and M. Dijkstra, Phys. Rev. E, {\bf 68}, 021407 (2003).

\bibitem{kleinetalJPCM2002}
R. Klein, H.H. von Gr\"unberg, C. Bechinger, M. Brunner and V. Lobaskin
J. Phys.: Condens. Matter, {\bf 14}, 7631 (2002).

\bibitem{piel:melzer2002}
For a recent review see
A. Piel and A. Melzer, Adv. Space. Res., {\bf 29}, 1255 (2002), and references
therein.

\bibitem{Valuina}
O. Vaulina, S. Khrapak and G. Morfill, Phys. Rev. E ,{\bf 66}, 16404 (2002).

\bibitem{rosenfeldPRE1994}
Y. Rosenfeld, Phys. Rev. E, {\bf 49}, 4425 (1994).

\bibitem{farouki:hamaguchi1994}
R.T. Farouki and S. Hamaguchi, J. Chem. Phys., {\bf 101}, 9885 (1994).

\bibitem{Stillinger4}
F.H. Stillinger and T.A. Weber, Phys. Rev. B, {\bf 22}, 3790 (1980).

\bibitem{paper3}
F.H. Stillinger and D.K. Stillinger, Physica A, {\bf 244}, 358 (1997).

\bibitem{paper5}
A.A. Louis and P.G. Bolhuis and J-P. Hansen, Phys. Rev. E, {\bf 62},
7961 (2000).

\bibitem{paper4a}
A. Lang and C.N. Likos and M. Watzlawek and H. L{\"o}wen, J. Phys.: Condens.
Matter, {\bf 12}, 5087 (2000).

\bibitem{PrestipinoetalPRE2005}
S. Prestipino, F. Saija, P.V. Giaquinta, Phys. Rev. E {\bf 71},
050102(R) (2005).

\bibitem{IngoJCP}
I.O. G\"otze, H.M. Harreis and C. N. Likos, J. Chem. Phys. {\bf 120},
7761 (2004).

\bibitem{LairdKrollPRA1990}
B.B. Laird and D.M. Kroll, Phys. Rev. A {\bf 42}, 4810 (1990).

\bibitem{McConnellGastPRE1996}
G.A. McConnell and A.P. Gast, Phys. Rev. E {\bf 54}, 5447 (1996).

\bibitem{GrohSchmidtJCP2001}
B. Groh and M. Schmidt, J. Chem. Phys {\bf 114}, 5450 (2001).

\bibitem{BausJCP1990}
M. Baus, J. Phys.: Condens. Matter, {\bf 2}, 2111 (1990).

\bibitem{footnote}
Note that as $\alpha \rightarrow 0$ Eq.\ (\ref{eq:F_ideal_crystal}) is no longer
a good approximation to Eq.\ (\ref{eq:F_id}) with $\rho(\rr)$ given by Eq.\
(\ref{eq:rho_solid}) -- i.e.\ the overlap between the density peaks on
neighbouring lattice sites becomes significant. In practice, this occurs in
regions of the phase diagram where the system is a fluid. However, one must be
aware that as $\alpha \rightarrow 0$, Eq.\ (\ref{eq:F_ideal_crystal}) gives a
spurious divergence $\rightarrow -\infty$ in the free energy.

\bibitem{Lindemann}
F. Lindemann, Z. Phys., {\bf 11}, 609 (1910).

\bibitem{VondeLageandBethePR1947}
F.C. Von der lage and H.A. Bethe, Phys. Rev., {\bf 71}, 612 (1947).

\bibitem{TarazonaPRL2000}
P. Tarazona, Phys. Rev. Lett., {\bf 84}, 694 (2000).

\bibitem{footnote1}
This is due to the fact that the DFT with exess Helmholtz free energy given by
Eq.~(\ref{eq:F_ex_2}) reduces to the lattice sum. Hence, the fcc/bcc transition
is driven by the duality relations of the GCM, first noticed by Stillinger --
see Refs.~\cite{Likos,paper3,StillingerPRB1979}.

\bibitem{StillingerPRB1979}
F.H. Stillinger, Phys. Rev. B, {\bf 20}, 299 (1979).

\bibitem{footnote2}
When the difference in coexisting densities is small, the conditions for
coexistence between phases of equal pressures and chemical potentials in the two
phases become equivalent to equating the free energies $F$ in the two phases.

\bibitem{myThesis}
A.J. Archer, {\it Statistical Mechanics of Soft Core Fluid Mixtures},
PhD thesis, University of Bristol, (2003).

\bibitem{WatzlaweketalPRL1999}
M. Watzlawek, C.N. Likos, and H. L\"owen, Phys. Rev. Lett., {\bf 82}, 5289
(1999). 

\bibitem{HamaguchietalPRE1997}
S. Hamaguchi, R.T. Farouki and D.H.E. Dubin, Phys. Rev. E, {\bf 56}, 4671
(1997).

\bibitem{kremeretalPRL1986}
K. Kremer, M.O. Robbins, and G.S. Grest, Phys. Rev. Lett, {\bf 57}, 2694 (1986).

\bibitem{robbinsetalJCP1988}
M.O. Robbins, K. Kremer and G.S. Grest, J. Chem. Phys., {\bf 88}, 3286 (1988).

\bibitem{rosenberg:thirumalaiPRA1987}
R.O. Rosenberg and D. Thirumalai, Phys. Rev. A, {\bf 36} 5690 (1987).

\bibitem{meijer:frenkelJCP1991}
E.J. Meijer and D. Frenkel, J. Chem. Phys., {\bf 94}, 2269 (1991).

\bibitem{HamaguchietalJCP1996}
S. Hamaguchi, R.T. Farouki and D.H.E. Dubin, J. Chem. Phys., {\bf 105}, 7641
(1996).

\bibitem{Hoy:RobbinsPRE2004}
R.S. Hoy and M.O. Robbins, Phys. Rev. E, {\bf 69} 056103 (2004).

\bibitem{Hopkins1}
P. Hopkins, A.J. Archer and R. Evans, Phys. Rev. E, {\bf 71} 027401 (2005).

\bibitem{LikosetalJPCM1994} 
C.N. Likos, Zs.T. N\'emeth and H. L\"owen, J. Phys.: Condens.
Matter, {\bf 6}, 10965 (1994).

\bibitem{RasconetalJCP1997}
C. Rasc\'on, E. Velasco, L. Mederos and G. Navascu\'es, J. Chem. Phys.,
{\bf 106}, 6689 (1997).

\bibitem{Archer6}
A.J. Archer, C.N. Likos and R. Evans, J. Phys.: Cond. Matter, {\bf 16}, L297
(2004).

\bibitem{Pusey}
P.N. Pusey in {\it Liquids, Freezing and the Glass Transition},
Proceedings of the Les Houches Summer School, Session LI, 3-28 July 1989
Ed. J.-P. Hansen, D. Levesque and J. Zinn-Justin, part 2.

\bibitem{dendrimers}
C.N. Likos, S. Rosenfeldt, N. Dingenouts, M. Ballauff, P. Lindner, N. Werner,
and F. V\"ogtle, J. Chem. Phys., {\bf 117}, 1869 (2002).

\end{thebibliography}
\end{document}